# SCOUT: Prefetching for Latent Structure Following Queries

Farhan Tauheed[†‡], Thomas Heinis[†], Felix Schürmann[‡], Henry Markram[‡], Anastasia Ailamaki[†]

[†]*Data-Intensive Applications and Systems Lab,* [‡]*Brain Mind Institute*
*École Polytechnique Fédérale de Lausanne, Switzerland*
{firstname.lastname}@epfl.ch

## ABSTRACT

Today's scientists are quickly moving from in vitro to in silico experimentation: they no longer analyze natural phenomena in a petri dish, but instead they build models and simulate them. Managing and analyzing the massive amounts of data involved in simulations is a major task. Yet, they lack the tools to efficiently work with data of this size.

One problem many scientists share is the analysis of the massive spatial models they build. For several types of analysis they need to interactively follow the structures in the spatial model, e.g., the arterial tree, neuron fibers, etc., and issue range queries along the way. Each query takes long to execute, and the total time for executing a sequence of queries significantly delays data analysis. Prefetching the spatial data reduces the response time considerably, but known approaches do not prefetch with high accuracy.

We develop SCOUT, a *structure-aware* method for prefetching data along interactive spatial query sequences. SCOUT uses an approximate graph model of the structures involved in past queries and attempts to identify what particular structure the user follows. Our experiments with neuroscience data show that SCOUT prefetches with an accuracy from 71% to 92%, which translates to a speedup of 4x-15x. SCOUT also improves the prefetching accuracy on datasets from other scientific domains, such as medicine and biology.

## 1. INTRODUCTION

By moving from purely in vitro experiments to simulating models in silico, scientists in many different disciplines such as biology, physics, neuroscience, etc., have experienced a drastic paradigm shift in recent years. Analyzing a natural phenomenon, rebuilding it as a model and simulating it, often leads to a more profound understanding of the mechanisms behind it. The tools scientists use, however, typically involve inefficient data management techniques which complicate and delay the analysis of massive models.

To build and analyze spatial models, the scientists use spatial indexes that help them to efficiently retrieve precisely defined parts of the model by executing range and other spatial queries. Many such indexes [27, 21, 14] have been developed in recent years to provide fast random access to spatial data. Scientists, however, often need *navigational* access to the data, i.e., moving from one location to another one nearby, based on a certain logic. Support for efficient navigation through spatial data is still a challenge.

Physicians, bioinformaticians and neuroscientists use navigational access to analyze their data such as the bronchial tracks of the lung, the arterial tree or the neuron network extracted from CT scans and then converted into three dimensional representations. In their spatial datasets they follow a structure, for example lung tracks, artery, neuron branches etc., and issue range queries along the way. The locations of the queries are hence not random, but are guided by the path of interest of the scientist, correlating with structures in the spatial model.

In these cases navigational access to spatial data happens as *guided spatial query sequence*, i.e., as a sequence of $n$ three dimensional spatial range queries whose locations are determined by a guiding structure. A guiding structure is one out of the many spatial structures in the spatial dataset the user follows and hence it intersects with all of his $n$ range queries. The spatial structures in a dataset are the structures that provide the user with a navigational path and have a dataset specific semantic, e.g., neuron branches, artery and lung airway as shown in Figure 1. The range queries in the sequence are executed interactively; based on the current query result, the user decides where to go next. The range queries are adjacent to each other, slightly overlapping or with small gaps between them.

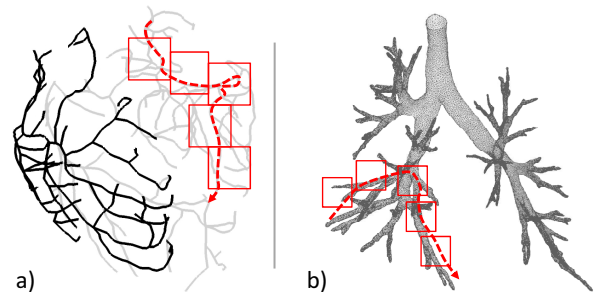

Figure 1: Guided spatial query sequences on a 3D model of (a) a pig's heart arterial tree and (b) a human lung airway track with gaps between the queries in the sequence. The dotted line indicates the guiding structure.

Due to the interactive nature of the problem, the I/O subsystem is idle during the time that results are analyzed and during the time the user decides the location of the next range query. Clearly, prefetching is an excellent technique to improve performance of navigational access to spatial data, where data can be prefetched between two consecutive queries. The impact of prefetching is of course minute if the entire dataset fits into main memory (or SSD or similar low latency storage media) or in case the path chosen is known a priori and where prefetching is hence perfect. Scientific datasets, however, have been growing rapidly and many still do not fit into main memory. Similarly, in many settings, e.g., interactive applications, the path that the user chooses cannot be known beforehand. For many of today's applications based on spatial data, prefetching is thus still an important method to speed up queries. However, as we will show, current prefetching approaches for spatial data do not perform well, because they only rely on previous query positions.

We therefore develop SCOUT, which instead of using the position of past queries like other approaches do, considers the previous query content and identifies the guiding structure among the many spatial structures in the range query results. It achieves this by using an approximate graph representation of the spatial objects and traverses it to predict the location of the next query. By using a graph representation, SCOUT is independent of the complexity or geometry of the dataset. SCOUT additionally uses heuristics to reduce the cost of prediction by building sparser graphs in subsequent spatial queries. Several optimizations are also presented that improve prefetch accuracy and speed up computing the prediction when SCOUT is used with a particular class of spatial indexes.

## 2. RELATED WORK

Prefetching is a popular technique to hide data transfer costs, e.g., in client-server communication [24], or when loading instructions [16] and data [6] into CPU caches.

Prefetching along graph structures is also used in different contexts. Web pages prefetching [20] uses a page-link graph to predict the user's next request (links followed by many users or similar learning mechanisms are used). For object oriented databases [12] and data structures [25] like binary trees and linked lists, tables containing all pairs of vertices in the graph representation are precomputed. When navigating through the graph, the tables are used to quickly locate the next vertex. While all these approaches help to find the next vertices in the graph given the current vertex, they are orthogonal to the problem we address as SCOUT targets at finding the set of vertices (the guiding structure) the user is following.

Research in locating spatial data has focused heavily on designing indexes [10]. Prefetching spatial data is orthogonal to spatial indexes, except when an index is used to improve prefetching performance. More particular to our problem, however, are the following approaches used to prefetch spatial data. We categorize them based on what information is used to predict the future location as static methods, trajectory extrapolation methods and learning approaches.

### 2.1 Static Methods

Static methods use heuristics for predicting the future query location and do not consider any past information. The *Layered* [31] approach segments the spatial data into a grid and prefetches all surrounding grid cells for future queries. The *Hilbert-Prefetch* [22] works similarly but assigns each cell a Hilbert value (based on the cell's coordinates) and prefetches cells with similar Hilbert values like the one of the current location.

Instead of the application level, the multimap approach [19] works at the disk level and ensures that data close in space is physically stored close together on disk. When reading entire disk pages from the disk, the I/O subsystem automatically prefetches data around the location queried for and hence this method is similar to the Layered approach.

### 2.2 Trajectory Extrapolation Methods

Trajectory extrapolation methods assume that navigational access to spatial data follows a path. They use the past query locations, interpolate them with a polynomial and extrapolate the new location. While the *Straight Line Extrapolation* [26] uses the last two query positions and a simple linear extrapolation, the *Polynomial* approach [4, 5] uses several previous query positions (more than two) as well as a polynomial of degree two (or higher) to extrapolate the next query location. For visualization applications, the *Velocity* [30] approach additionally uses the user's velocity on the motion path to determine the next query location.

Finally, the *EWMA* [7] approach uses exponential weighted moving averages to assign each past movement vector of the query a weight, adds up all vectors and extrapolates the future movement. The parameter $\lambda$ controls the weight assigned to previous queries: the last query is weighted with $\lambda$, the second to last with $(1-\lambda) * \lambda$, and so on.

### 2.3 Learning Approaches

User behavior can also be learned and be used to prefetch data in case users always chose to explore the spatial data along similar paths or guiding structures. The *sequence pattern mining* [9] approach collects past user behavior and mines it (with a clustering algorithm) to anticipate future user behavior. Other data mining or association rules learning approaches [2, 8, 28] use apriori algorithms. Different learning techniques based on *Bayesian network* [29] and *Markov chains* [13] have also been proposed.

For massive models, however, learning from past user behavior does not significantly improve prediction accuracy. Because the models are huge, they contain virtually infinitely many possible paths, reducing the probability that sufficient users take the same paths.

## 3. MOTIVATION

Our work is motivated by the needs of scientists who face significant performance challenges when querying spatial data in general, and the Blue Brain Project (BBP [17]) we collaborate with in particular. In the following we first discuss the neuroscience application, then describe the use cases in which the neuroscientists' work can be sped up by prefetching spatial data and finally illustrate the opportunity for prefetching spatial data. With a set of experiments we demonstrate the limited accuracy of known approaches and motivate the need for new prefetching approaches.

### 3.1 Neuroscience Use Cases

The quest for understanding the human brain, urges the neuroscientists in the BBP to build and simulate models of the brain at an unprecedented scale. In their models, each neuron is represented with several hundred of three-dimensional cylinders, modeling the structure of the neuron soma and the branches extending from the soma, bifurcating several times. Today they build models with 500,000 neurons represented by 2.3 billion cylinders amounting to 165 GB on disk. The amount of data will grow substantially in the future as the ultimate goal of the Blue Brain Project is to simulate a model of the human brain which will contain nearly 86 billion neurons and will require several hundred Petabytes of disk space.

The neuroscientists in the BBP need to work with guided spatial query sequences in many different aspects of their work. Although using state-of-the-art spatial indexes speeds up access to the spatial data considerably, random reads in spatial indexes throughout spatial query sequences create a bottleneck and slow down the analysis, inadvertently affecting the productivity of the neuroscientists. Efficient support of sequences of range queries is therefore key and has substantial impact on the quality of the neuroscientists' research. Guided spatial query sequences are pivotal in the following use cases:

**Ad-hoc Queries:** These queries are used to correct errors in the models introduced in the imaging process. They follow the branches of the neuron, execute a range query at every step and carry out an analysis of the query result. The type of analysis depends on the error to be rectified. In some cases, tissue statistics need to be calculated, in others, structural pattern matching algorithms need to be executed.

**Model Building:** To place synapses, i.e., the elements connecting the neurons, neuroscientists need to follow some of the branches and detect where their proximity to another branch falls below a given threshold. Calculating the exact distance between two branches is a computationally expensive problem.

**Walkthrough Visualization:** The three-dimensional walkthrough visualization is primarily used for discovering structural anomalies. The neuroscientist follows the neuron branch and executes range queries for visualization.

While the query pattern in these use cases is always a guided spatial query sequence, the use cases differ in parameters such as volume of the queries, sequence length and so on.

### 3.2 Prefetch Opportunity

In the use cases described the user executes a query and waits for the results. Once the result is retrieved, the user analyzes the results and then issues the next query in the sequence. The disk, however, remains idle during the time results are being analyzed and the time can hence be used to prefetch results of future queries in the sequence. We refer to this time as the *prefetch window*. Ideally all data needed to answer the next query should be read from disk during the prefetch window. However, because the prediction is not always entirely accurate, *residual I/O* is needed to retrieve the results not prefetched.

Figure 2 illustrates the timeline of resources utilized by three spatial queries in a sequence. The first query is executed and at the end of it the prediction computation is started. While the user analyzes the first query result, the prediction is used to prefetch data into a cache. Upon execution of the next query by the user, as much data as can be found is retrieved from the cache. The cache misses/residual I/O is subsequently retrieved directly from disk.

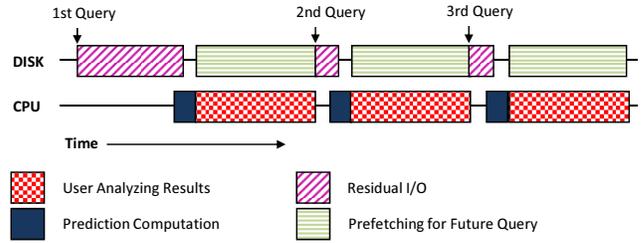

Figure 2: Resource usage timeline for guided spatial query sequences.

### 3.3 Accuracy of the State-of-the-Art

We demonstrate the limitations of existing prefetching approaches by measuring the prediction accuracy for query sequences executed on a 450 million cylinder model from our collaboration with neuroscientists. We use synthetic query sequences so we can more easily vary the volume of each single query. The sequences have a length of 25 queries (a complete description of the experimental setup follows in Section 7.1).

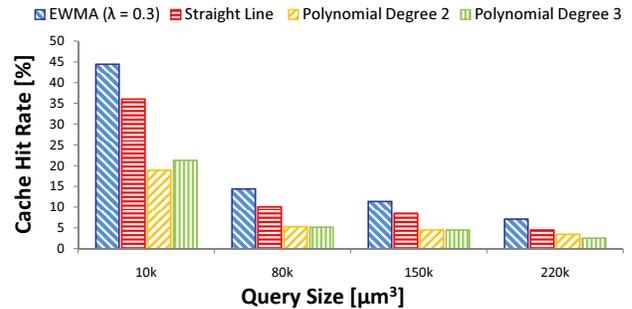

Figure 3: Prediction accuracy of state-of-the-art approaches.

We compare the best approaches from Section 2, i.e., *EWMA* (with the $\lambda$ that yields best accuracy) as well as the *Polynomial* interpolation (with varying degrees and using as many recent query locations to interpolate as their degree plus one) and measure the prediction accuracy as the cache hit rate[1]. The results as a function of the volume of the query are shown in Figure 3.

Interpolating the query positions with the *Polynomial* approaches is not a good approximation of the neuron structure the scientist is following, as even the best Polynomial interpolation approach never exceeds a prediction accuracy of 37%. Furthermore, because polynomials with a higher degree tend to oscillate, the accuracy drops with an increasing degree. Weighting the previous query positions and giving the most recent ones more weight improves the accuracy as the results of *EWMA* show. Still, even in its best configuration ($\lambda = 0.3$), EWMA's hit rate does not exceed 44%.

A common problem of all approaches is that the prediction accuracy drops substantially as the volume of query increases. The accuracy drops because in large queries there is a higher probability that the structure being followed bifurcates or bends, leading to a jagged query trace that cannot be interpolated well. Clearly only relying on previous query positions is not enough to reliably predict the future position of the query. In this paper we demonstrate that the content of past queries must also be considered to improve prediction accuracy.

---

[1]Percentage of data read from the prefetch cache rather than from disk.

### 3.4 SCOUT Overview

As we demonstrated with the motivation experiments, taking into account only the locations of previous queries is not enough to predict future query locations with high enough accuracy. To support scientists in executing guided spatial range query sequences efficiently, SCOUT therefore not only considers the locations of previous queries, but also their content. While the result of query $q$ in the sequence is loaded, SCOUT already starts to reconstruct the structures in $q$ and approximates them with a graph. Once the graph is constructed, it is traversed to find the locations where it exits $q$. At the exit locations, the graph (and not the query locations) is extrapolated linearly to predict the next query locations. By also considering the content of previous queries, SCOUT achieves a significantly higher prefetch accuracy.

## 4. PREDICTION

SCOUT accesses the spatial data through a spatial index and stores the spatial data in the prefetch cache. Any spatial index can be used as long as it can execute spatial range queries. User requests are served out of the prefetch cache and are loaded into it in case of cache misses. To speed up the prediction process, building the graph and retrieving the range query result is interleaved: while the result is read, the graph is already assembled.

To predict the location of the next query, SCOUT uses the content of the last queries and reduces the structures to a graph. Each range query result contains several structures, making it very hard to predict the next query location. By considering the result of several past sequential range queries, SCOUT can reduce the number of structures the user may be following and can predict the next location.

In the following we discuss how to find guiding structure information in spatial datasets, how SCOUT represents this guiding structure information efficiently as a graph, how it reduces the candidate set of possible structures the user is following and finally, how the candidate set is used to predict future query locations.

### 4.1 Guiding Structure Information

In many spatial datasets the guiding structures, i.e., the structures the user may follow, are explicit. For example in the case of datasets like road networks [15], models of the lung [1] or the arterial tree [11] the guiding structure are explicitly represented by spatial structures which define the routes in road networks, bronchial tracks and arteries respectively. SCOUT can directly use explicit representations of guiding structure information to build a graph.

In other datasets the guiding structure, however, is not explicit but implicit. In the case of earthquake simulation based on unstructured meshes for ground models, for example, the dataset contains interesting features like seismic fault lines the user may follow. The guiding structures in this example, however, are not explicitly but implicitly represented by regions of high density in the meshes.

SCOUT cannot directly use implicit guiding structures directly and a preprocessing step is required to make the guiding structures explicit. The transformation can be done manually through analysis or algorithms like medial axis transformation [18, 23] can be used to identify the topological skeleton which oftentimes correlates with the guiding structure. The particular method chosen to identify the implicit structure depends on the dataset (as well as its semantics) and is outside the scope of SCOUT.

### 4.2 Graph Representation

With an explicit guiding structure in the dataset, SCOUT can predict the next query location by summarizing the spatial strucures in the query result as a graph representation. Many scientific applications already use spatial datasets with an underlying graph structure, where vertices of a graph are represented by spatial objects and edges represent two objects connected with each other. Lung airway track models are an excellent example where the spatial model is represented with polygon meshes (as is common in many graphics and GIS applications). Given the common representation of polygon meshes as a list of polygon vertices and a list of polygons faces (the latter referencing the former), SCOUT can easily extract a graph with vertices represented by polygon faces and edges connecting adjacent polygon faces.

In case no underlying graph representation is available, SCOUT builds one on the fly. An example is the pig's heart arterial tree where arteries are represented using three-dimensional objects(cylinders) which, unlike polygon meshes, do not have adjacency information. Graphs are built based on spatial proximity, i.e., where objects are represented by vertices and spatially close objects are connected with edges. Given a set of $n$ spatial objects to build the graph a brute force algorithm requires $O(n^2)$ pairwise proximity comparisons of objects. To speed up the graph building process SCOUT uses spatial grid hashing. Grid hashing partitions the entire three-dimensional space of range query into equi-volume grid cells and each object is mapped to grid cells based on how many grid cells it intersects with. Finding the intersection between an arbitrary geometry object and the grid cell can be time consuming and we thus use any one of the three most commonly used geometry simplification techniques. A minimum bounding rectangle surrounding the object, a straight line or a point can be used depending on the geometry of the object. For example in our arterial tree dataset we approximate the cylindrical object by a straight line as illustrated in Figure 4. Objects mapped to the same grid cells are then connected with edges in the graph representation. The guiding structure is one of the many paths in this graph representation.

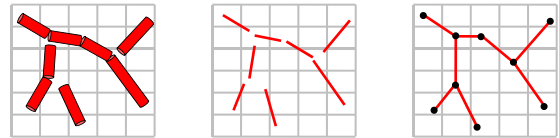

Figure 4: Building the approximate graph by mapping objects to grid cells and connecting them if they occupy the same cell.

Grid hashing is controlled by setting the resolution, i.e., volume of the grid cells. If the resolution is too coarse, many spatially distant objects map to the same grid cell, thereby increasing the number of edges in the resulting graph. Too coarse a resolution also increases the time needed to create the graph, because all objects in the same cell need to be connected pairwise. More importantly, predictions become less accurate because a coarse resolution leads to more edges than the brute force technique. Excess edges can imply structures that are not present in the dataset and may mislead the prediction process. If, on the other hand, the resolution is too fine, objects that are close together and should be connected, end up in different cells. As a result, the graph is sparser(fewer edges) when compared to the one obtained through the brute force technique. The graph traversal in the prediction process will run into many dead ends and hence the prediction accuracy will be lower. Our strategy is to use a fine resolution and work with sparser approximate graph representation. In fact, the constructed graph does not need to be entirely precise, because it is used as a hint for future query locations. As we show in Section 7.4.5, accurate predictions can be obtained with approximate graphs.

### 4.3 Iterative Candidate Pruning

In every query result there are many different spatial structures and the user may be following any one of them, making the prediction of the location of the next query very difficult. To identify the structure the user follows, SCOUT exploits the fact that all queries in the guided spatial range query sequence must contain the structure followed. SCOUT inspects the two recent query results to identify the set of structures $x$ that exit the $(n-1)^{th}$ query and the set of structures $e$ that enter the $n^{th}$ (the most recent) query. The intersection $x$ and $e$ is the candidate set and contains the guiding structure the user follows. By repeating this process for subsequent queries in the sequence, candidates can be pruned, effectively reducing the set of candidates as shown in Figure 5.

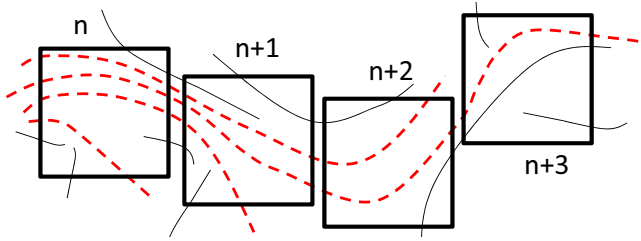

Figure 5: Pruning the irrelevant structures (solid lines) from the candidate set (dashed lines) in subsequent queries (solid squares) of the sequence.

With many queries in a sequence pruning reduces the candidate set considerably. Already after only a few queries, the one structure followed can be identified. In case of the ad-hoc queries use case, the structure followed is oftentimes identified after six queries. In case of a *reset*, i.e., when the user decides to abandon following a particular spatial structure in favour of another, the candidate set again contains all spatial structures from the last range query result.

### 4.4 Predicting Future Query Locations

Based on the candidate set and the graph of all structures in the $n^{th}$ query SCOUT predicts the future query locations. It starts at the edges representing the structures in the candidate set and traverses the graph depth first to find the locations where the graph exits the query. Performing this traversal is efficient and is linear in terms of number of edges and vertices of the graph representation. Pruning the candidate set further reduces the cost of traversing the graph and it decreases with the increase in number of range queries in the sequence.

SCOUT uses the edges exiting the current query and extrapolates them linearly to predict the locations of the next queries. More complex approaches like extrapolation with higher order curves do not yield better results.

## 5. PREFETCHING

We have so far only discussed the ideal case where subsequent queries are adjacent without gaps in-between and where the duration of the prefetch window is known. In practice neither of these assumptions holds and in the following we discuss how to adapt the prefetching strategy.

### 5.1 Prefetch Window Duration

Estimating the duration $DW$ of the prefetch window, i.e., how long the computation (visualizing the query result or others) takes and how long the user needs to determine the next query location is difficult. Underestimating $DW$ means we do not make the most of the prefetch window and waste resources; overestimating it means we will not be able to fully execute the query at the predicted location. To complicate matters, $DW$ can vary between queries.

Instead of estimating $DW$, SCOUT uses an incremental prefetch technique which stops once the user issues the next range query in the sequence. It is important that in the limited time window SCOUT prefetches data closer to the exit location of the candidate structure $E$ with higher priority because prefetching data far away from $E$ is more likely to be prefetched unnecessarily. This strategy is illustrated in Figure 6.

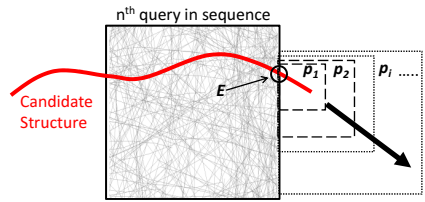

Figure 6: Incremental prefetching queries $p_i$ (dashed lines) are executed along the extrapolated axis starting at $E$.

To implement this strategy we require to read the data in a particular spatial order from disk, i.e., data close to $E$ first. Only few spatial indexes, however, permit to retrieve the query result in a particular order. To work with arbitrary spatial indexes, SCOUT executes a prefetch query $p_i$ with a small region $r_i$ to prefetch data from the predicted location. It then iteratively increases the volume of each subsequent prefetching query such that the region $r_{i+1}$ of the next prefetching query $p_{i+1}$ is bigger. The center of the prefetching queries is shifted along the extrapolated axis of the candidate structure, until the prefetch window duration ends.

Because the intersection of $r_i$ and $r_{i+1}$ is not empty, some of the content of $p_i$ is retrieved again when executing $p_{i+1}$. Data is, however, not read repeatedly from disk because the result of $p_i$ is still in the cache and hence this operation incurs only little or no overhead.

## 5.2 Prefetching Strategies

Predicting the future query location is simple if there is only one structure in the candidate set. Oftentimes the candidate set, however, is bigger because it is either a) the first query of the sequence and SCOUT has not yet reduced the candidate set or b) the structure the user follows bifurcates into two or several structures. We use two strategies to deal with multiple candidate structures.

### 5.2.1 Deep Prefetching

An aggressive strategy picks any one structure at random out of the candidate set $C$ and uses it for predicting the next query location as shown in Figure 7 (a). The entire prefetch window duration is used to retrieve $D$ data from this location. Intuitively, this strategy predicts correctly with a probability $\frac{1}{|C|}$ (assuming disjoint prefetch locations). On average $\frac{D}{|C|}$ data is correctly prefetched but the prefetch accuracy varies widely, resulting in varying execution times.

While SCOUT does not use any application specific information, it is nevertheless possible to improve results by using application specific heuristics that allows picking the correct structure out of the candidate set $C$, thereby improving the average accuracy of the deep prefetching strategy. For example in the context road networks, the user may issue query sequences to retrieve data along the road with the highest speed limit. If the dataset contains semantic information about speed limits, it can be used to pick the one structure from the candidate set.

### 5.2.2 Broad Prefetching

Broad prefetching is a defensive strategy that gives all of the structures in the candidate set $C$ equal weight and prefetches at all predicted locations proportionally as is shown in Figure 7 (b). Assuming a prefetch window during which $D$ data is prefetched, $\frac{D}{|C|}$ will be prefetched at each location. Because the next query will be executed at one of these locations, $\frac{D}{|C|}$ is prefetched correctly for each query individually and on average. Prefetching at several locations thus does not increase prefetching accuracy, but the variation in prediction accuracy decreases.

The queries at several predicted locations may overlap because their respective exit locations are close. Instead of retrieving the overlapping content several times (from cache or disk), the regions $R_1$ and $R_2$ of two overlapping queries are expanded until the area of $R_1 \cup R_2$ is equal to the sum of the areas $R_1$ and $R_2$.

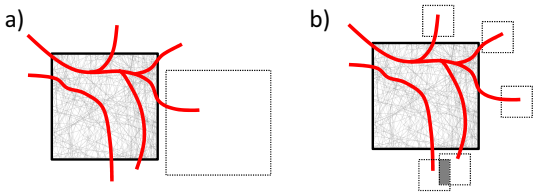

Figure 7: Deep (a) and Broad Prefetching (b).

Clearly, executing several small prefetching queries instead of only one big query incurs additional overhead. For a small number of queries the overhead is insignificant but for several thousand small queries the overhead may become substantial. In this case it is necessary to limit the number of structures considered for prefetching.

If the number of prefetching queries is limited to a number $d$, their locations should be chosen so that areas where many candidate structures exit the query are prefetched. We use a k-means approach [3] to find $d$ clusters and, to ensure that we prefetch around an exit location of a candidate, choose an exit location at random in each cluster. Because k-means has a smoothed polynomial complexity, it does not impose an undue overhead for our application.

## 5.3 Gaps Between Queries

In some use cases, the queries are not adjacent to each other but have gaps between them. In the case of visualization, scenes with a distance between them are rendered to create the illusion of high speed movement. For our neuroscience use case, for example, the scientist wants to analyze regions along the same neuron branch with gap intervals to ensure the isolation of the local electrical field.

To support these scenarios we use linear extrapolation from the exit locations of the guiding structures and we estimate the gap distance. The gaps between queries can be of any size, but they are typically governed by a particular characteristic of the use case, e.g., velocity of visualization, and remain the same throughout a sequence of queries. We thus use the distance between the last two queries as a prediction for the next gap.

Linear extrapolation predicts the next query location accurately in case of small gaps. The larger the gap, however, the more likely is the guiding structure to twist and bend after exiting the recent query and the less likely it becomes that a linear extrapolation will approximate the structure. For larger gaps the prediction accuracy thus drops. As we show in Section 6.3 the prediction accuracy can be improved considerably by using a particular class of spatial indexes.

## 6. OPTIMIZATIONS

We design SCOUT to work with any spatial access method/index. If, however, it is possible to use an index that a) allows the retrieval of pages from disk in a particular spatial order and b) stores the relative positions of objects (neighborhood information), then we can optimize graph building as well as prediction in case of gaps.

### 6.1 Ordered Retrieval of Spatial Data

Both FLAT [27] and DLS [21] are example indexes that allow for the ordered retrieval of spatial data. The indexes work by maintaining neighborhood information for each object. Both indexes execute queries using a two phase approach. First, they find an arbitrary object located inside the query region and then, using the neighborhood information, recursively visit the neighbors until no more results inside the query region can be found.

While DLS works mainly for convex tetrahedral meshes by using the neighborhood information already present in the dataset, FLAT on the other hand works on arbitrary spatial datasets by first computing the spatial neighborhood information as a pre-processing step. Using this information the retrieval order of the pages can be controlled. This, however, is not possible with traditional indexes such as variants of R-tree [14], because they lack neighborhood information. In the following we discuss how we can exploit this ordered retrieval to reduce the time for building the graph and how to improve the prediction accuracy in case of gaps.

### 6.2 Sparse Graph Construction

SCOUT builds the graph and traverses it starting from the edges representing the candidate set to the locations where they exit the query region. We interleave the process of building the graph with retrieving the result, i.e., each page that is part of the result is added to the graph once it is read from disk. Traversing the graph to determine the exit locations, however, can only be done once the graph is complete, i.e., when the entire result is retrieved.

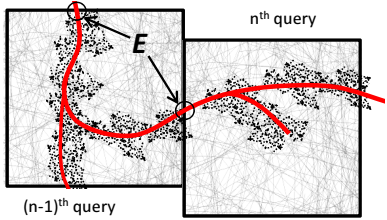

Figure 8: Construction of sparse graph (dotted) using only the relevant pages around the candidate structure (solid curve).

We can start the traversal earlier if we use indexes where we control the order in which the result pages are retrieved. We retrieve the pages we need for building the graph first and traverse the graph while we load the remaining pages of the result. More particularly, we retrieve the pages at the exit locations of the previous query first, build the subgraphs of each page $P$, start to traverse the subgraph and find the locations $X$ where the subgraph exits the page $P$. Subsequently, we recursively repeat this process and retrieve all neighboring pages of $P$ at $X$.

With this mechanism we ensure that we only retrieve those pages that contain objects reachable from the exit locations $E$ of the previous query as shown in Figure 8. Once all the pages containing the parts of the graph that are reachable from $E$ have been retrieved, the remaining pages that are relevant for the user but irrelevant for the prediction are read. While the remaining pages are being read, the prediction process can already start to traverse the graph to determine the new exit locations and hence to predict the new query locations.

This approach has two benefits: first, it reduces the cost for building and traversing the graph and second, the prediction process is already finished once the query result is retrieved and prefetching can start immediately.

### 6.3 Gap Traversal

Linear extrapolation for predicting the next query location works well for small gaps, but fails to give good prediction accuracy for bigger gaps. The prediction accuracy is lower because for a big gap the guiding structure is more likely to bifurcate and bend, making it hard to approximate it correctly with a linear extrapolation.

By using a spatial index like FLAT [27] or DLS [21], we can extract exactly those pages needed to follow the structures that exit the current query. We use a similar approach like the one used in Section 6.2 for sparse graph construction but we continue the traversal outside the query region. Starting from the exit locations of the last query, we load exactly those pages $P$ that neighbor the exit locations and build the subgraph of $P$'s content. We traverse the subgraphs of all $P$, identify the exit locations $E$ and load the neighboring pages at $E$. The process is repeated recursively to load exactly those pages needed to reconstruct the graph outside the query region, as illustrated in Figure 9. The traversal ends once the estimated gap distance is reached.

The grey grid in Figure 9 represents the data pages containing spatial objects. The shaded grid cells illustrate the pages visited in the gap region. Here the guiding structure is a neuron fiber which can bifurcate and change its trajectory.

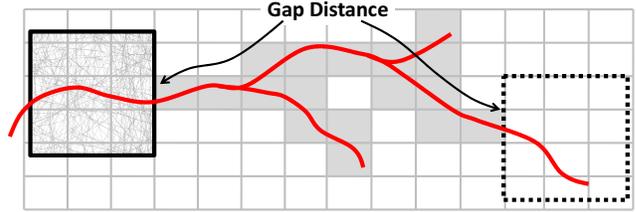

Figure 9: Traversing pages in the gap region using neuron fiber as a guiding structure.

The gap traversal technique requires that we retrieve data between queries that is not required by the user and we hence trade additional I/O for a better prediction accuracy. To minimize the overhead (which can be substantial for large gaps) we set the *gap I/O budget*, a maximum allowable limit of pages retrieved in the gap. In case no extra I/O can be spent for bridging the gap, we resort to a backup mechanism, e.g., linear extrapolation from the point where the traversal was stopped.

## 7. EXPERIMENTAL EVALUATION

In this section we first describe the experimental setup and demonstrate the benefit of SCOUT by benchmarking it with microbenchmarks designed with the use cases of the BBP as a basis. We also study the performance of SCOUT with a sensitivity analysis.

### 7.1 Setup and Methodology

The experiments are run on a Linux Ubuntu 2.6 machine equipped with 2 quad CPUs AMD Opteron, 64-bit @ 2700MHz and 16GB RAM. The storage consists of 4 SAS disks of 300GB capacity each, striped to a total of 1TB.

For the measurements we use a brain tissue model with a volume of $285mm^3$ containing 100,000 neurons. The neurons are modeled with 450 million three-dimensional cylinders amounting to 33GB on disk. Each cylinder is described by two end points and a radius for each endpoint. With 79% of the 33GB dataset, the majority of the data is used to describe the geometry (the cylinders) while the remaining 21% are used to describe additional attributes like the identifiers and type of the spatial objects. The additional information is, however, not used by SCOUT as we do not exploit any application specific information. For building the graph, SCOUT reduces the cylinder to a line segment by solely using the two endpoints.

We use two versions of prefetching algorithms for experiments. The simple version $SCOUT$ is coupled with widely used R-Tree(STR Bulkloaded) [14] spatial index for accessing data. The optimized version $SCOUT\text{-}OPT^2$ implements the techniques described in Section 6 and therefore must be coupled with FLAT [27]. We compare these algorithms against the *Straight Line Extrapolation*, *EWMA $\lambda=0.3$* and *Hilbert Prefetch* approaches as described in Sections 2 and 3.

Spatial indexes have tunable configuration parameters. For both indexes we use a 4KB page size and a fanout of

---

[2]SCOUT-OPT is only used in experiments which contain gaps in the query sequences. In the absence of gaps SCOUT and SCOUT-OPT have the same performance.

87 objects per page, both indexes are bulk loaded using a fill factor of 100%. These parameters can affect the performance of the index and thus indirectly the performance of prefetching with SCOUT. Tuning spatial indexes is, however, outside the scope of this paper.

We allow 4GB of memory to cache prefetched data. We leave the remaining 12GB of RAM for the user to perform his analysis. After executing each sequence of queries, we clear the prefetch cache, the operating system cache and the disk buffers.

## 7.2 Microbenchmark Design

In the following we define the microbenchmarks used in the experiments. Our microbenchmarks are designed based on query templates used in the real use cases and have different values for each different parameter, i.e., volume of the query, sequence length and gap distances as set by the neuroscientists. We use 30 sequences for all the benchmarks, all other parameters are summarized in Figure 10.

We also use input from the neuroscientists to set the prefetch window duration. If $d$ time is required to retrieve the data from disk for one query of a particular use case, and it takes $u$ time to process the data, then we define the prefetch window ratio as $r = u/d$. Therefore, if $(0 < r \leq 1)$ then the use case is I/O bound, whereas if $(r > 1)$ the use case is CPU bound. This ratio is used to simulate the effect of different algorithms processing the data in the use cases. SCOUT, however, does not estimate the prefetch window duration but instead uses the incremental prefetching discussed in Section 5.1.

### 7.2.1 Ad-hoc Queries

Neuroscientists frequently execute queries to correct errors in the models introduced while scanning. Their algorithms first find the spatial objects belonging to a neuron of interest and then validate them against many predetermined test cases. The tests performed at every step depend on the type of error to be rectified, in some cases tissue statistics need to be calculated while in others structural pattern matching algorithms need to be executed.

We define two ad-hoc queries benchmarks that differ primarily in the prefetch window duration. Tests based on statistical analysis are generally computationally cheap and therefore we set the prefetch window ratio to 0.8. For pattern matching, on the other hand, algorithms are more time consuming and we set the prefetch window ratio to 1.4.

### 7.2.2 Model Building

To place synapses neuroscientists follow some of the branches and detect where their proximity to another branch falls below a given threshold. Computing the exact distance between two branches is computationally expensive and therefore there is a big window of opportunity to prefetch data (prefetch window ratio of 2). We design a model building benchmark with rather a small query volume of $20,000 \mu m^3$ and the query sequence typically consists of 35 spatial queries.

### 7.2.3 Walkthrough Visualization

The three-dimensional walkthrough visualization is primarily used to monitor fiber reconstruction and to help in discovering structural anomalies. In this use case, a series of view frustum culling[3] operations is required along the neuron branch the neuroscientist follows. The frustum culling directly translates into a sequence of spatial queries with a volume (enclosing the view frustum) of $30,000 \mu m^3$. We design two walkthrough visualization benchmarks: a first one inspired by a low quality, but fast polygon rendering algorithm which uses a 1.2 prefetch window ratio and a second based on a more time consuming ray tracing algorithm (better quality) which uses a 1.6 prefetch window ratio.

In some cases, users issue query sequences with gaps, for example, to increase the speed of the walkthrough visualization. We design two additional different benchmarks for visualization with gaps, with similar parameters as the aforementioned visualization benchmarks, but with a gap distance as described in Figure 10.

|  | Queries in Sequence [# of Queries] | Query Volume [μm³] | Aspect Ratio | Gap Distance [μm] | Prefetch Window [ratio] |
|---|---|---|---|---|---|
| Ad-hoc Queries (Stat. Analysis) | 25 | 80K | Cube | 0 | 0.8 |
| Ad-hoc Queries (Pattern Matching) | 25 | 80K | Cube | 0 | 1.4 |
| Model Building | 35 | 20K | Cube | 0 | 2 |
| Visualization (Low Quality) | 65 | 30K | Frustum | 0 | 1.2 |
| Visualization (High Quality) | 65 | 30K | Frustum | 0 | 1.6 |
| Visualization with Gaps (High Quality) | 65 | 30K | Frustum | 25 | 1.2 |
| Visualization with Gaps (Low Quality) | 65 | 30K | Frustum | 25 | 1.6 |

Figure 10: The parameters of the microbenchmarks.

## 7.3 Experimental Results

We compare SCOUT to related approaches for executing the microbenchmarks described in Section 7.2. SCOUT is compared against the best variants of the related approaches: Straight Line Extrapolation approach, EWMA 0.3 and Hilbert prefetching.

In a first experiment we use the benchmarks without gaps and measure the accuracy of prediction as the cache hit rate. The results are shown in Figure 11(a). Clearly, SCOUT outperforms the other approaches, even exceeding an accuracy of 90%. In the model building benchmark the longer prefetch window duration helps in increasing the prediction accuracy. The reason is that more spatially close data is cached with the incremental prefetch strategy. The visualization benchmarks on the other hand have a shorter prefetch window, but longer sequences of queries. This allows SCOUT to reduce the candidate set effectively by using iterative candidate pruning and hence helps to detect the guiding structure with higher probability. SCOUT, however, is comparatively less accurate for the ad-hoc query benchmarks because a) the query sequences are shorter, b) the prefetch window is shorter and c) the volume of the queries is bigger (which makes it likelier that the guiding structure bifurcates). Comparing the results of the two ad-hoc queries benchmarks clearly shows that with a larger prefetch window (bigger opportunity), the prefetching accuracy also increases. The corresponding difference in accuracy when running the two visualization benchmarks is not as significant. In the second experiment we measure the speedup of the query response time when using a prefetching approach compared to not using prefetching at all. The results are shown in Figure 11(b) and correlate with the accuracy.

---

[3]The view frustum is the volume that contains everything that is potentially (there may be occlusions) visible on the screen. Culling is the process of extracting the spatial data that is enclosed inside the view frustum.

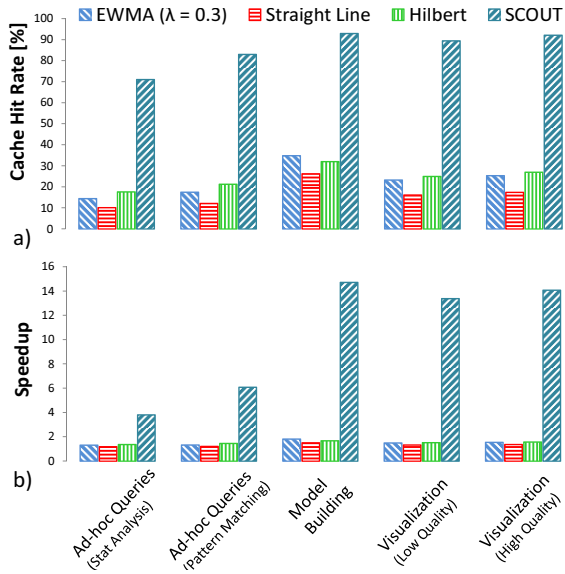

Figure 11: (a) Accuracy of the approaches for all microbenchmarks. (b) Speedup of the approaches for all microbenchmarks (compared to no prefetching).

When running benchmarks with gaps between queries the accuracy of SCOUT is slightly higher than the accuracy of trajectory extrapolation techniques. The reason is that SCOUT, in case of gaps also uses a straight line extrapolation, in addition to traversing the candidate structures. SCOUT-OPT, on the other hand, performs much better because it uses FLAT to traverse the gap by following the candidate structures and only retrieves the data needed. Figure 12 (right) shows the speedup of query sequence response time. Although SCOUT is slightly more accurate than trajectory extrapolation techniques, it does not yield higher speedup, because with gaps the prediction step becomes a substantial overhead. Indeed, SCOUT has no way to prune candidates in the gap region and is forced to traverse the entire graph for each query in the sequence.

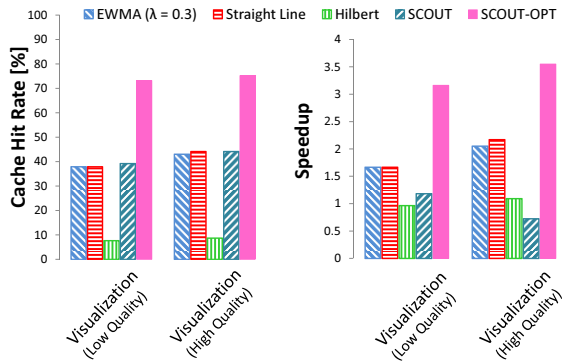

Figure 12: Accuracy and speedup of benchmark with gaps between queries.

## 7.4 Prediction Accuracy Sensitivity Analysis

In the following experiments we test how SCOUT's accuracy (and the speedup therewith) depends on the different parameters of the query sequence. Unless stated otherwise we experiment with a neuroscience dataset containing 450 million elements executing for each experiment 50 sequences of 25 queries, each query having volume of 80,000 $\mu m^3$ and a prefetch window ratio of 1. Because the performance of SCOUT-OPT is the same as that of SCOUT in the absence of gaps, we only measure the performance of SCOUT-OPT for benchmarks with gaps.

### 7.4.1 Volume of Query

To measure the effect of increasing the volume of the query, we repeat the experiments by increasing the volume of the query by 35,000 $\mu m^3$. The results in Figure 13 (a) show that the accuracy drops gradually with the increase in volume of the query: the larger the volume, the more likely is it that the candidate structures bifurcate, resulting in more branches that the user may be following and thereby making it harder to detect the one structure the user is following. The speedup drops accordingly from 9 to 4.5.

### 7.4.2 Dataset Density

We measure the effect of increasing the dataset size by increasing the number of neurons (or spatial objects it is modeled with). For each experiment we increase the size and along with it spatial density by adding 50 million more objects in the volume of 285 $mm^3$, which allows us to test if SCOUT will scale to more dense, future models. As Figure 13 (b) shows, despite the increased data density, the accuracy stays at around 80% while the speedup remains constant (around 5.5). As data density increases, more data must be retrieved, but also, the user analysis takes longer, translating into a proportionally longer prefetch window.

### 7.4.3 Sequence Length

The effect of increasing the sequence length is shown in Figure 13 (c) where we increase the number of queries in each sequence by 10. Longer query sequences increase the prediction accuracy substantially, reaching 93.1%. As we argued previously, by using the iterative candidate pruning mechanism, the longer a query sequence, the smaller the candidate set, and hence the better SCOUT can predict the future query location. The speedup increases accordingly from 7x to 20x.

### 7.4.4 Prefetch Window

By increasing the prefetch window ratio we allow more time for prefetching. As shown in Figure 13 (d), the accuracy increases from 29% to 88% as the ratio is increased from 0.1 to 2.5, illustrating that SCOUT becomes more effective for computationally intense use cases. The opposite also holds true: when SCOUT is used with I/O bound use cases the prefetch opportunity decreases, resulting in a lower prefetch accuracy because less data can be prefetched. Interestingly, varying the prefetch window has the same effect as varying the prefetch cache size. A small sized cache will halt prefetching prematurely once it becomes full, leading to lower accuracy. Similarly, accuracy increases for bigger cache sizes as it does with a longer prefetch window.

### 7.4.5 Graph Precision

The grid hashing mechanism helps to efficiently build an approximate graph representation of the spatial dataset as described in Section 4.2. In a next experiment we measure the effect of the grid resolution on the prefetch accuracy. To make the grid resolution coarser we decrease the number of grid cells by a factor of eight in every experiment. As the results in Figure 13 (e) show, even with only 512 grid cells the approximate graph delivers good prediction accuracy, but drops substantially once the resolution is decreased further.

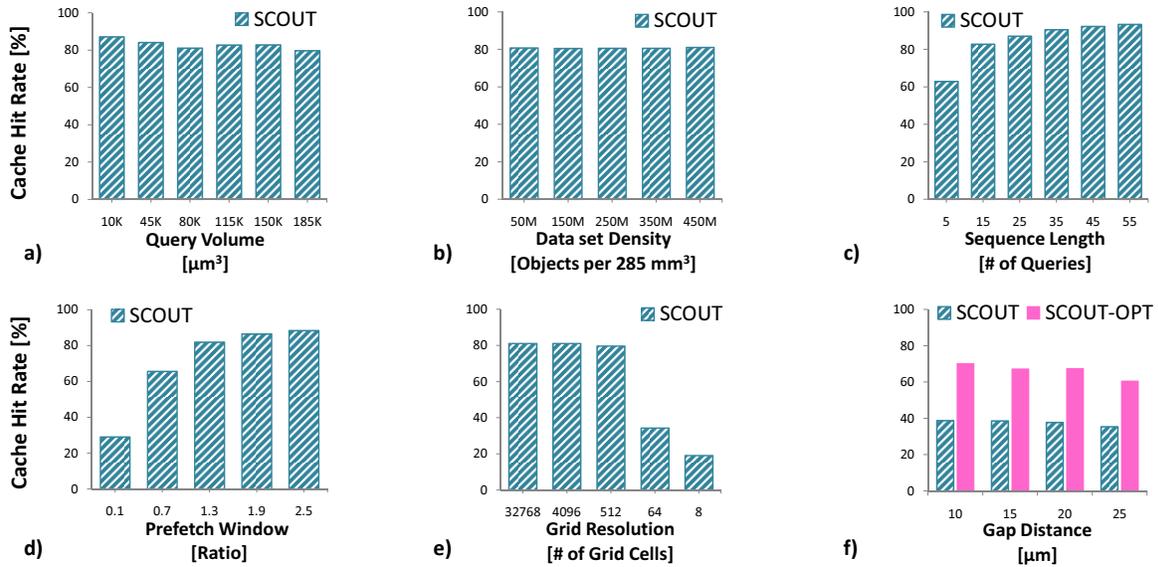

Figure 13: Sensitivity analysis of prediction accuracy.

### 7.4.6 Gap Distance

We increase the gap distance between the queries of a sequence and measure the effect on prediction accuracy. We test the impact on both SCOUT and SCOUT-OPT. The results in Figure 13 (f) show that an increase in the gap distance has a negative effect on the prediction accuracy. SCOUT-OPT performs much better because of the gap traversal technique (Section 6.3). However, the accuracy decreases as the gap distance is increased because longer gaps require more I/O. We use a fixed I/O budget of 10% of the pages used in the recent query to ensure that the gap traversal does not become an overhead.

## 8. SCOUT ANALYSIS

In the following subsections we analyze SCOUT by testing it with several synthetic queries. We vary the characteristics of the query sequences and quantify the cost of prediction.

### 8.1 Query Response Time Breakdown

To understand the overhead of SCOUT in the face of evermore dense, future datasets, we break down the query response time into graph building time, prediction (graph traversal) and residual I/Os (time needed to retrieve data due to misprediction) while increasing the dataset density. With an increase in data density, also the result size grows.

The results in Figure 14 show that the time needed to build the graph remains around 15% of the total time and hence using optimizations such as grid based hashing enable building the graph on the fly instead of precomputing it. The prediction time based on the graph traversal takes up to 6% of the time. The trend also indicates that there is no increase in the relative cost of modeling and prediction as the result size grows.

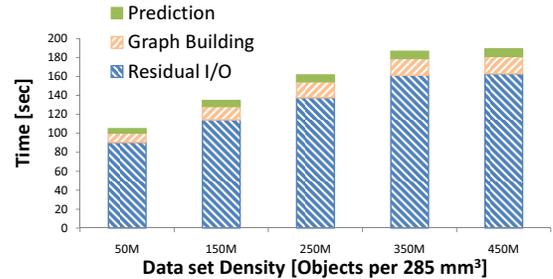

Figure 14: SCOUT time breakdown for graph building time, prediction time and time for retrieving remaining data after prefetching.

### 8.2 Graph Building Cost

The graph building time depends on the number of spatial objects in the query region. To understand the relationship between the result set size and the graph building time we execute 35 sequences containing 25 queries each. For the measurement, we add up the graph building time for all 25 queries in the sequence and plot it against the total results obtained for the 25 queries.

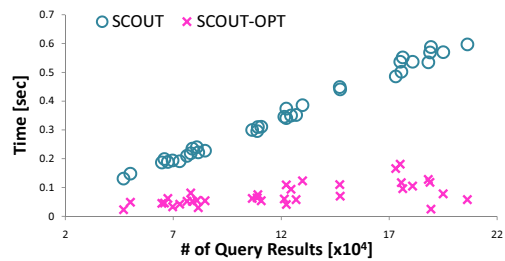

Figure 15: Modeling time relative to the number of objects in result.

As the results in Figure 15 show, the graph building for SCOUT depends linearly on the number of results in the query. The relation is linear because we use all of the spa-

1540

tial objects in the result set to construct the graph. For SCOUT-OPT the graph building time scales even better than SCOUT because the graph built only uses spatial objects belonging to candidates structures, using the *sparse graph construction* approach.

We also measure the memory needed to store the major data structures, i.e., the graph (adjacency list) and queues used for graph traversal, required by SCOUT and SCOUT-OPT. The memory required relative to the space needed for the query results is nearly 24% for SCOUT and because it only builds the subgraph needed for prediction, only 6% for SCOUT-OPT.

### 8.3 Prediction Cost

The iterative candidate pruning technique reduces the time to traverse the graph structure in subsequent queries because it reduces the candidate set and thus only the subgraph reachable from the edges representing the candidate set needs to be traversed.

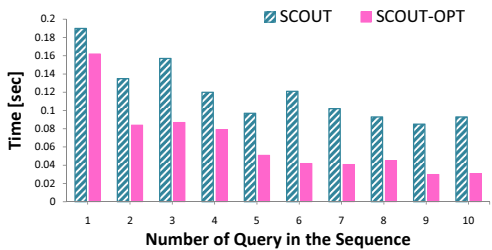

Figure 16: Graph traversal operations required for a sequence of 10 spatial queries.

We illustrate this with an experiment where we use 50 sequences with 10 queries each and measure the time taken for prediction divided by the number of elements in the result of each query. Figure 16 shows that the prediction time per result element of each query indeed decreases the further we progress in the sequence. Because it uses *sparse graph construction*, SCOUT-OPT generally takes less time for the prediction/graph traversal.

### 8.4 Applicability

Applications of guided spatial query sequences are frequent in scientific applications, for instance for analyzing kidney models, water pollution studies using a river network, animal herd migration studies using terrain models, spatial analysis of epidemiological models etc. Other than our neuroscience application, we test SCOUT on an arterial tree model of the pig's heart [11]. The dataset contains 2.1 million 3D cylinders (154MB on disk). We also use human lung airway models [1] containing 7.1 million (527MB on disk) triangular surface meshes shown in Figure 1.

SCOUT infers the guiding structure rather than relying on the user to provide this information in the application layer. In fact in the development of SCOUT we do not make any application or dataset specific assumptions and hence, it can potentially be used in other non-scientific domains. An example use case is fetching spatial data in proximity to a road network route from a mobile device. In this case no time consuming analysis is required on the results, however, data can still be prefetched during the time the user needs to make a decision what route to follow. Consequently, accurate prefetching becomes key for effectively using the limited prefetch memory available on the device. We therefore use a road network of North America [15] modeled with 7.2 million two-dimensional line segments (531MB on disk).

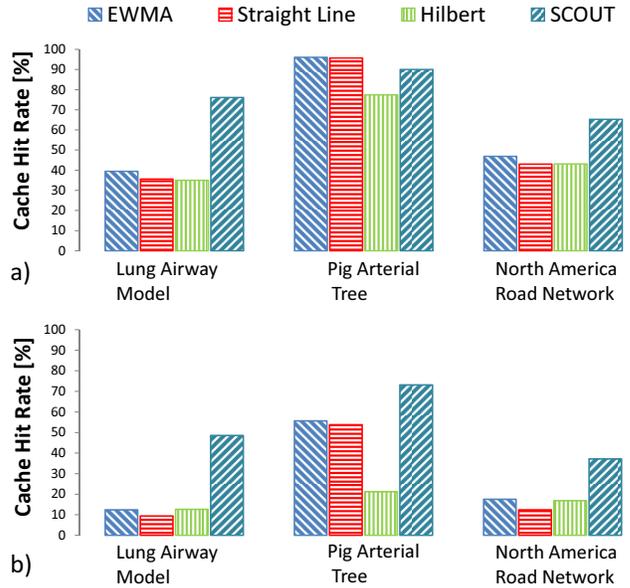

Figure 17: Prediction accuracy comparison for various spatial datasets using (a) small (b) large volume queries.

For each dataset we compare the prediction accuracy by executing 50 sequences of two different volume of queries, small volume queries are $5 \times 10^{-7}$ smaller than the dataset volume and large volume queries are $5 \times 10^{-4}$ times smaller. Each sequence contains 25 spatial range queries and the queries are generated by making a random walk on the graph representation of the entire dataset. We use a prefetch window ratio of one.

As shown in Figure 17 (a), compared to other approaches SCOUT performs better for the lung airway model and road networks. For the arterial tree, however, trajectory extrapolation approaches provide higher accuracy, up to 96% (EWMA), while SCOUT only achieves a prediction accuracy of up to 90.1%. SCOUT does not perform well because the dataset consists of arterial branches, i.e., smooth structures that can be interpolated very well with polynomials (or the weighted movement vectors of EWMA). For large queries, on the other hand, SCOUT performs better in all cases and predicts with up to 73% accuracy as is shown in Figure 17 (b). For the arterial tree SCOUT performs better because with the increase of volume of query the arterial branches are more likely to bifurcate and bend, making it more difficult to interpolate them with polynomials.

### 8.5 Limitations

Curve extrapolation techniques can outperform SCOUT in situations where the query sequence can be well approximated by straight lines or polynomials as the experiments in Figure 17(a) (in Section 8.4) with the arterial tree dataset and small volume queries show. This is typically the case when the dataset contains simple structures that are closely followed with small queries. Even in this case, however, SCOUT prefetches with more than 90% accuracy. In case the queries have larger volumes the query trace becomes more jagged, they can no longer be approximated well with curves and SCOUT therefore clearly outperforms curve extrapolation approaches.

Although SCOUT oftentimes identifies the correct structure after only few queries in the sequence, the guiding structure can bifurcate, making the identification of the structure followed more difficult. Application specific heuristics can be used for selecting the branches. However, for the sake of general applicability, we refrain from using any such technique. After all, SCOUT's high accuracy of 75% to 92% in our experiments leaves little room for improvement.

## 9. CONCLUSIONS

Scientists frequently use guided spatial query sequences to analyze their vast spatial datasets. Prefetching the spatial data is an excellent approach to speed up the query sequences, state-of-the-art prefetching approaches for spatial data, however, do not predict future query locations accurately. To speed up the execution of the query sequences we develop SCOUT, a novel prefetching approach that considerably departs from the state of the art, because it does not only consider the last query positions, but uses the content of the last queries to predict the future query position.

SCOUT exploits that the user does not move through the data randomly, but typically follows a structure. It reliably identifies the structure followed (or at least it reduces the candidate set considerably) by summarizing the structures to a graph representation. SCOUT prefetches with an accuracy of up to 92% and achieves a speed up of up to 15× on neuroscience workloads.

Because we do not make any particular assumptions about the characteristics of spatial datasets when designing SCOUT, it can also be used in different applications and domains. With experiments we show that it considerably speeds up navigational access to datasets from other domains as well.

## 10. REFERENCES

[1] T. Achenbach et al. Accuracy of automatic airway morphometry in computed tomography-correlation of pathological findings. *European Journal of Radiology*, 81(1):183 – 188, 2010.
[2] R. Agrawal and R. Srikant. Mining sequential patterns. In *ICDE*, pages 3–14, 1995.
[3] D. Arthur et al. K-means has polynomial smoothed complexity. In *FOCS*, pages 405–414, 2009.
[4] R. Azuma and G. Bishop. A frequency-domain analysis of head-motion prediction. In *SIGGRAPH*, pages 401–408, 1995.
[5] A. Chan, R. Lau, and A. Si. A motion prediction method for mouse-based navigation. In *CGI*, pages 139 –146, 2001.
[6] S. Chen, A. Ailamaki, P. Gibbons, and T. Mowry. Improving hash join performance through prefetching. In *ICDE*, pages 116 – 127, 2004.
[7] J. Chim et al. On caching and prefetching of virtual objects in distributed virtual environments. In *MULTIMEDIA*, pages 171–180, 1998.
[8] J. Choi, S. H. Noh, S. L. Min, and Y. Cho. Towards application file-level characterization of block references: A case for fine-grained buffer management. In *SIGMETRICS*, pages 286–295, 2000.
[9] T. T. Fu, S.-S. Hung, H.-L. Lin, D. Tsaih, and J. T. Chen. Intelligent-based latency reduction in 3d walkthrough. In *ISTASC*, pages 218–226, 2010.
[10] V. Gaede et al. Multidimensional access methods. *ACM Computing Surveys*, 30(2):170–231, 1998.
[11] L. Grinberg et al. Large-scale simulation of the human arterial tree. *Clinical and Experimental Pharmacology and Physiology*, 36(2):194–205, 2009.
[12] N. Knafla. A prefetching technique for object-oriented databases. In *BNCOD*, pages 154–168, 1997.
[13] D. Lee et al. Adaptation of a neighbor selection markov chain for prefetching tiled web gis data. In *ADVIS*, pages 213–222, 2002.
[14] S. T. Leutenegger, M. A. Lopez, and J. Edgington. STR: a Simple and Efficient Algorithm for R-tree Packing. In *ICDE*, pages 497–506, 1997.
[15] F. Li et al. On trip planning queries in spatial databases. In *SSTD*, pages 273–290, 2005.
[16] C.-K. Luk and T. C. Mowry. Compiler-based prefetching for recursive data structures. In *ASPLOS*, pages 222–233, 1996.
[17] H. Markram. The blue brain project. *Nature Reviews Neuroscience*, 7(2):153–160, 2006.
[18] M. McAllister and J. Snoeyink. Medial axis generalization of river networks. *Cartography and Geographic Information Science*, 27(2):129–138, 2000.
[19] S. Minglong et al. Multimap: Preserving disk locality for multidimensional datasets. In *ICDE*, pages 926–935, 2007.
[20] A. Nanopoulos, D. Katsaros, and Y. Manolopoulos. A data mining algorithm for generalized web prefetching. *IEEE Transaction on Knowledge and Data Engineering*, 15(5):1155–1169, 2003.
[21] S. Papadomanolakis et al. Efficient query processing on unstructured tetrahedral meshes. In *SIGMOD*, pages 551–562, 2006.
[22] D.-J. Park and H.-J. Kim. Prefetch policies for large objects in a web-enabled gis application. *Data and Knowledge Engineering*, 37(1):65–84, 2001.
[23] E. Perlman, R. Burns, M. Kazhdan, R. R. Murphy, W. P. Ball, and N. Amenta. Organization of data in non-convex spatial domains. In *SSDBM*, pages 342–359, 2010.
[24] M. Rabinovich and O. Spatschek. Web caching and replication. *SIGMOD Record*, 32(4):107–108, 2002.
[25] A. Roth, A. Moshovos, and G. S. Sohi. Dependence based prefetching for linked data structures. In *ASPLOS*, pages 115–126, 1998.
[26] E. G. Said, E. B. Omar, and L. Robert. Data prefetching algorithm in mobile environments. *European Journal of Scientific Research*, 28(3):478–491, 2009.
[27] F. Tauheed et al. Accelerating range queries for brain simulations. In *ICDE*, pages 218–230, 2012.
[28] E. Thereska, M. Abd-El-Malek, J. J. Wylie, D. Narayanan, and G. R. Ganger. Informed data distribution selection in a self-predicting storage system. In *ICAC '06*, pages 187–198, 2006.
[29] A. Walker et al. A bayesian framework for automated dataset retrieval in geographic information systems. In *MMM*, pages 138–150, 2004.
[30] Z. Xiaohong, F. Shengzhong, and F. Jianping. A history-based motion prediction method for mouse-based navigation in 3d digital city. In *GCC*, pages 686 –690, 2008.
[31] J. Zhang and S. You. Dynamic tiled map services: Supporting query visualization of large-scale raster geospatial data. In *COM.Geo*, pages 191–198, 2010.